\documentclass[a4paper,11pt]{article}
\usepackage{pos}
\usepackage{amsmath}
\usepackage[export]{adjustbox}
\usepackage{float}

\title{Non-perturbative Renormalization of the EMT in Full QCD}

\author*[a]{Pavan}
\author[a]{Olaf Kaczmarek}
\author[b]{Guy D. Moore}
\author[a]{Christian Schmidt}

\affiliation[a]{Fakult\"at f\"ur Physik, Universit\"at Bielefeld, D-33615 Bielefeld, Germany}
\affiliation[b]{Institut für Kernphysik, Technische Universität Darmstadt, D-64289 Darmstadt, Germany}

\abstract{
The energy-momentum tensor (EMT) is the conserved current corresponding to space-time translation symmetry.
Its applications are remarkably diverse, ranging from the thermodynamics to the calculation of transport coefficients.
While the EMT is well-defined in the continuum up to a total derivative, with its coefficients fixed by Ward identities, its extension to lattice QCD is not straightforward. The primary challenge arises from the breaking of continuous space-time symmetries by the discrete lattice regulator.
Although the EMT can be constructed on the lattice in a way that yields the correct continuum limit, the operators are not uniquely defined.
In this proceeding, we construct the EMT for both pure-gauge theory and full QCD, discussing its renormalization in the specific context of determining the coefficients required for shear viscosity.
In this context, we present a comparative analysis of the trace anomaly, number density, pressure, energy density and enthalpy density with imaginary chemical potential for multiple $\beta$ values at approximately the same temperature, aimed for the continuum limit.
}

\FullConference{The 42nd International Symposium on Lattice Field Theory (LATTICE2025)\\
2 November - 8 November 2025\\
Mumbai, India\\}

\begin{document}
\maketitle
\section{Introduction}
The energy-momentum tensor (EMT) is the conserved current associated with space-time translation symmetry. 
In the hydrodynamic limit, the first-order dissipative corrections to the equilibrium EMT involve the shear ($\eta$) and bulk ($\zeta$) viscosities, which characterize the dissipative properties of the medium:
\begin{equation}
        T_{ij} - T^{eq}_{ij} = -\eta \, \left( \nabla_{i} u_{j} + \nabla_{j} u_{i} - \frac{2}{3} \, \delta_{ij} \, \nabla_{k} u_{k} \right) - \zeta \, \delta_{ij} \, \nabla_{k} u_{k}.
\end{equation}
Hydrodynamics serves as an effective field theory for the long-wavelength or low-frequency regime.
In this framework, these transport coefficients represent the integrated-out microscopic degrees of freedom.
These coefficients are very useful for understanding the quark-gluon plasma (QGP).
However, their determination remains challenging \cite{GM}: experimental measurements are often subject to significant systematic uncertainties, while perturbative expansions suffer from poor convergence \cite{JG}. Therefore, we rely on lattice QCD for a non-perturbative determination, despite its own inherent challenges, such as EMT renormalization and the complexities of spectral reconstruction.
While the shear viscosity has been successfully determined in pure-gauge theory on the lattice \cite{LA, HM, FW}, its extension to (2+1)-flavor QCD remains an unresolved challenge.

The lattice determination of the shear viscosity requires the calculation of Euclidean two-point correlation functions of the EMT, followed by spectral reconstruction. In continuum $SU(3)$ theory, the EMT operators under $SO(4)$ rotational symmetry are defined as \cite{SC}
\begin{align}
    T_{\mu \nu} (x) \equiv & \frac{1}{g_0^2} \left[F_{\mu \alpha} (x) F_{\nu \alpha} (x) - \frac{1}{4} \delta_{\mu \nu} F_{\rho \sigma} (x) F_{\rho \sigma} (x) \right] 
    \nonumber \\
  & + \frac{1}{4} \bar{\psi}(x) \left( \gamma_\mu \overleftrightarrow{D}_\nu + \gamma_\nu \overleftrightarrow{D}_\mu - \frac{1}{2} \delta_{\mu\nu} \gamma_\alpha \overleftrightarrow{D}_\alpha \right) \psi(x) - g_{\mu\nu} m \bar \psi \psi\,.
\end{align}
While the coefficients in the continuum are strictly constrained by Ward identities, the lattice regulator preserves only discrete translational and rotational symmetries, which are subgroups of the continuum $SO(4)$ group. Consequently, there is no unique or direct operator construction for the EMT on the lattice. To address this, we construct the EMT using gradient flow.

The lack of continuous symmetry on the lattice necessitates associating a distinct renormalization coefficient with each irreducible representation (irrep), as these can no longer be determined solely by Ward identities. We utilize the enthalpy density as a physical constraint to fix these coefficients. While this procedure is relatively straightforward in pure-gauge theory, the inclusion of dynamical fermions introduces significant complexity. In this proceeding, we provide details of our renormalization method in (2+1)-flavor QCD.

\section{Theoretical Background}
The renormalization of the EMT in (2+1)-flavor QCD represents a significant technical challenge; therefore, we first outline the procedure for pure-gauge theory to establish a conceptual foundation. While the underlying principles remain consistent, the inclusion of dynamical fermions introduces non-trivial complexities that are detailed later in this proceeding.

\subsection{Pure-glue Theory}
\begin{figure}[h!]
\centering
\includegraphics[width=0.475\textwidth]{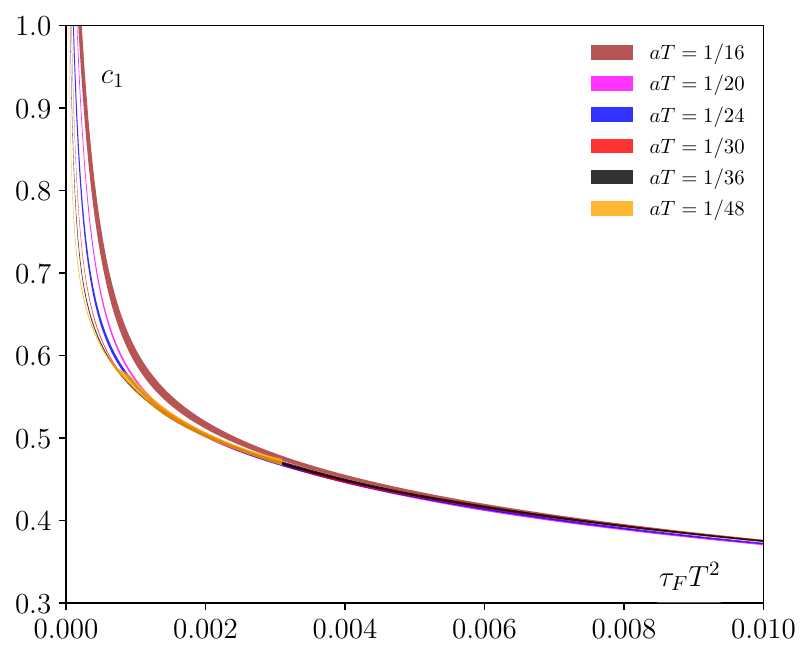}
\includegraphics[width=0.438\textwidth]{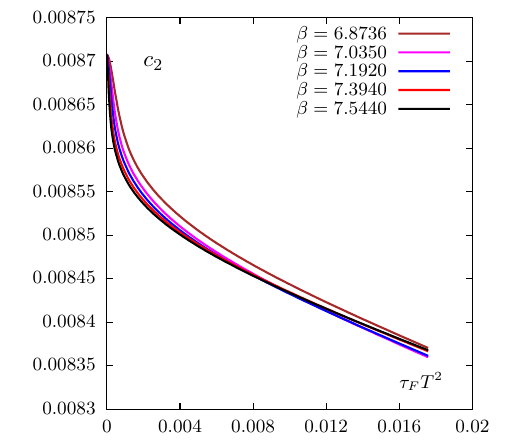}
\caption{The renormalization coefficient $c_1$ for various lattice spacings at fixed temperature (left). The renormalization coefficient $c_2$ measured at $T < T_c$ across different lattice spacings  (right). Plots from \cite{LA}.
\label{fig:1}}
\end{figure}

The EMT can be decomposed into two irreps under the $SO(4)$ group.
In the gradient flow formalism, the continuum $SO(4)$ symmetry on the lattice is justified by the smearing of operators over a space-time region of radius $\sqrt{8 \tau_F} > a$. Consequently, we expect discretization effects to be of order $\mathcal{O}(a^2/\tau_F)$. The flowed EMT operators in this formalism are \cite{HS}
 \begin{align}
   T^1_{\mu\nu}(x, \tau_F) &\equiv c_1(\tau_F) \left [ F_{\mu\rho}^a(x, \tau_F)F_{\nu\rho}^a(x, \tau_F) - \frac{1}{4} \delta_{\mu \nu} F_{\rho \sigma}^a(x, \tau_F)F_{\rho \sigma}^a(x, \tau_F) \right ]
\\
   T^2_{\mu\nu}(x, \tau_F) &\equiv c_2(\tau_F) \,
   \delta_{\mu\nu}F_{\rho\sigma}^a(x, \tau_F)F_{\rho\sigma}^a(x, \tau_F).
\end{align}
Determining the pressure or energy density independently would result in a system with one equation and two unknowns ($c_1$ and $c_2$). To resolve this, we adopt the methodology originally developed by Giusti and Pepe \cite{GP, GP2, GP3, GP4}.
In this framework, utilizing the enthalpy density ($\epsilon + p$) is a strategic choice for determining $c_1$, as the contributions from the pure trace term $T_{\mu \nu}^2$ cancel out:
\begin{align}
    \epsilon+p  &=  \left\langle   \frac{1}{3} \sum_{i = 1}^3 T_{ii}(\tau_F)- T_{00}(\tau_F)\right\rangle \nonumber \\
    &= c_1(\tau_F) \left \langle \frac{1}{3}  \sum_{i = 1}^3 F_{i \alpha} F_{i \alpha} -  F_{0 \alpha} F_{0 \alpha} \right \rangle
\end{align}
Furthermore, $c_2$ is determined via the scale dependence of the running coupling. As illustrated in Fig.~\ref{fig:1}, the renormalization coefficients $c_1$ and $c_2$ exhibit divergent behavior in the zero-flow-time limit ($\tau_F \rightarrow 0$). The gradient flow effectively suppresses these UV divergences, rendering the flowed quantities finite at finite flow time \cite{ML}. This is due to the fact that the gradient flow is a diffusion-like equation with a characteristic flow radius of $\sqrt{8 \tau_F}$. Since the products of these coefficients and the corresponding operators are associated to physical observables, the finiteness of the energy-momentum tensor is guaranteed when taking the $\tau_F \rightarrow 0$ limit.

This section provides a concise overview of the renormalization procedure in pure-gauge theory. For a comprehensive derivation and detailed discussion, we refer the reader to \cite{LA}.
\newpage
\subsection{Fermionic Theory}
The pure-gauge theory provides a useful template for the construction of the EMT in the presence of fermions. While the gluonic components remain identical, the inclusion of dynamical quarks introduce three additional irreps to the EMT under $SO(4)$ symmetry \cite{HH}:
  \begin{align}
   T^1_{\mu\nu}(x, \tau_F) &\equiv Z_1(\tau_F) \left [ F_{\mu\rho}^a(x, \tau_F)F_{\nu\rho}^a(x, \tau_F) - \frac{1}{4} \delta_{\mu \nu} F_{\rho \sigma}^a(x, \tau_F)F_{\rho \sigma}^a(x, \tau_F) \right ]
\\
   T^2_{\mu\nu}(x, \tau_F) &\equiv Z_2(\tau_F) \,
   \delta_{\mu\nu}F_{\rho\sigma}^a(x, \tau_F)F_{\rho\sigma}^a(x, \tau_F)
\\
   T^3_{\mu\nu}(x, \tau_F) &\equiv Z_3(\tau_F) \, \bar{\psi}(x) \frac{1}{2}
   \left [ \gamma_\mu\overleftrightarrow{D}_\nu
   +\gamma_\nu\overleftrightarrow{D}_\mu - \frac{1}{2} \delta_{\mu \nu} \gamma_\alpha\overleftrightarrow{D}_\alpha \right ]
   \psi(x)
\\
   T^4_{\mu\nu}(x, \tau_F) &\equiv Z_4(\tau_F) \,
   \delta_{\mu\nu}
   \bar{\psi}(x) \gamma_\alpha
   \overleftrightarrow{D}_\alpha
   \psi(x)
\\
   T^5_{\mu\nu}(x, \tau_F) &\equiv Z_5 (\tau_F) \,
   \delta_{\mu\nu}
   m_0 \bar{\psi}(x)
   \psi(x).
\end{align}
Among these five operators, only $T_{\mu \nu}^1$ and $T_{\mu \nu}^3$ contribute to the traceless components, which are the relevant operators for the determination of shear viscosity.
While the remaining operators are also required to construct the full correlator, the operators $T_{\mu \nu}^4$ and $T_{\mu \nu}^5$ are connected through the equations of motion; thus, the remaining associated coefficients can be fixed independently via the scale dependence of the running coupling and the quark masses (line of constant physics).
Consequently, the following discussion will focus specifically on the determination of $Z_1$ and $Z_3$.

We utilize the enthalpy density to fix these renormalization coefficients, following the same procedure established for the pure-gauge theory: 
\begin{align}
    \epsilon + p &= \left\langle \frac{1}{3} \sum_{i = 1}^3 T_{ii} - T_{00} \right\rangle \nonumber\\
    &= \left\langle Z_1(\tau_F)\, \left[ \frac{1}{3} \sum_{i = 1}^3 F_{i \alpha} F_{i \alpha} -  F_{0 \alpha} F_{0 \alpha} \right]  - Z_3(\tau_F)\, \bar{\psi} \left[ \frac{1}{3} \sum_{i = 1}^3 \gamma_i  \overleftrightarrow{D}_i - \gamma_0 \overleftrightarrow{D}_0  \right] \psi \right\rangle.
\label{ep_ferm}    
\end{align}
In contrast to the pure-gauge sector, the inclusion of dynamical fermions introduces an additional term, resulting in an underdetermined system with two unknowns for a single equation; therefore, independent information is required to uniquely determine these renormalization coefficients. To resolve this, we utilize measurements from two distinct ensembles to simultaneously extract both coefficients.
The procedure is detailed in the following section.

\section{Renormalization Method}
\begin{figure}[H]
\centering
\includegraphics[width=0.99\textwidth]{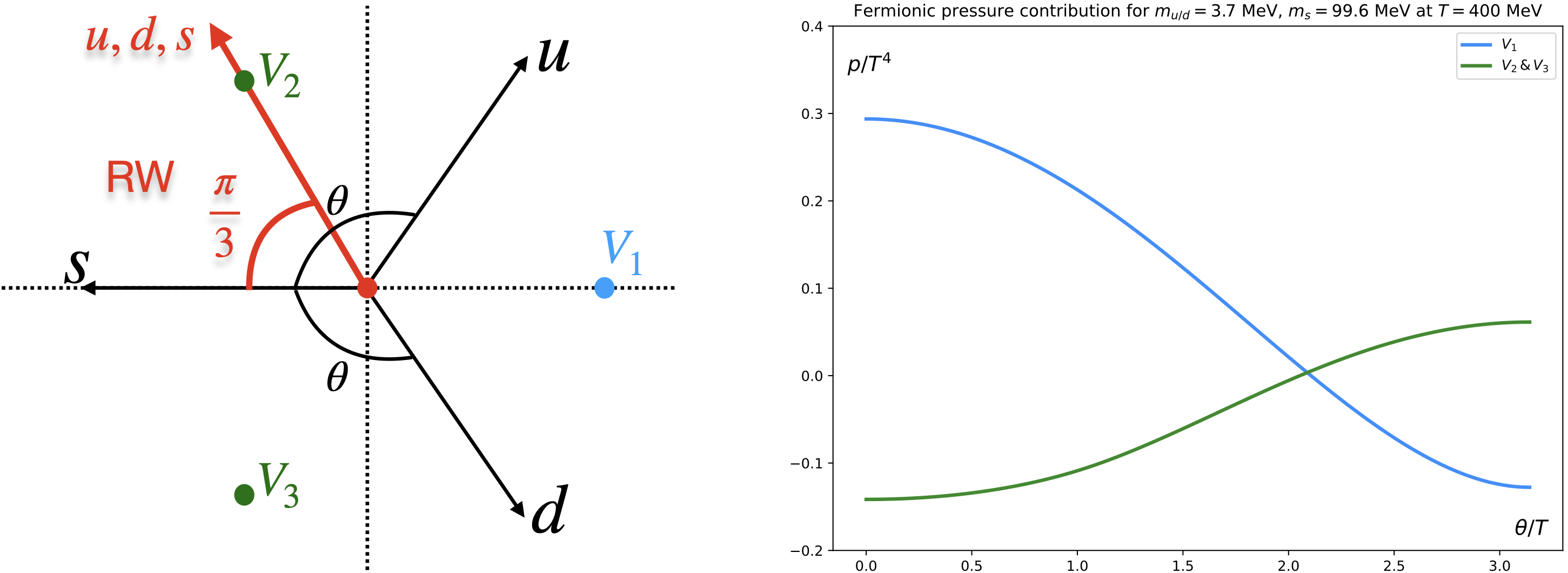}
\caption{Possible Polyakov loop vacua $V_{1,2,3}$, and the quark twist angles for the Roberge-Weiss case (red) and our case (black) (left). Free fermionic pressure for the Polyakov-loop vacua $V_1$, $V_2$ and $V_3$ (right). 
\label{fig:2}
}
\end{figure}

As demonstrated by Eq.~\ref{ep_ferm}, we have a system of two unknowns but only a single equation. To uniquely determine both renormalization coefficients, we utilize an imaginary chemical potential. The core strategy is to suppress the fermionic contribution in a second ensemble while keeping the gluonic component relatively unchanged. This approach is rooted in the $\mathbb{Z}_3$ center symmetry of QCD. In the pure-gauge limit, the three Polyakov loop vacua are degenerate and equally probable until they are broken spontaneously above a certain temperature. While the presence of fermions breaks this symmetry explicitly, the phase structure at finite temperature remains predominantly governed by center symmetry \cite{RW}. \\
The most prominent manifestation of this is the Roberge-Weiss (RW) transition. In the standard setup $(\mu_l/T = i \theta )$, a phase transition occurs at $\theta = \pi /3 $ as the system shifts between preferred vacua to minimize free energy. At this point, the fermionic contribution to the pressure is reduced by a factor of approximately two.
We utilize a modified configuration: $\mu_u / T =  -\mu_d /T = i \theta $ and $\mu_s = 0$. In this specific setup, the transition occurs at $\theta = 2 \pi /3 $ because the strange quark remains in the $V_1$ vacuum, stabilizing it for a larger range.
As shown in the right plot of Fig.~\ref{fig:2}, the pressure associated with vacuum $V_1$ dominates until $2 \pi /3$. At this transition point, the fermionic pressure is suppressed by a factor of 81 in a massless free theory. At physical quark masses, the ratio exceeds 100 but a heavier strange quark triggers the phase transition slightly earlier in free theory.
Since the gluons contribute a total of 16 degrees of freedom and fermions contribute 36 with a weight factor of 7/8, we expect the total pressure suppression to be around a factor of 3.

Motivated by the observations above, we perform our analysis for enthalpy density using the following ensembles:
\begin{enumerate}
    \item $\mu_u /T = \mu_d /T = \mu_s /T = 0 ~~(\equiv \mu_i)$ \label{ensemble1}
    \item $\mu_u/T  = -\mu_d/T  = i 2 \pi /3$ and $\mu_s/T = 0 ~~(\equiv \mu_f)$
    \label{ensemble2}
\end{enumerate}
We will perform our analysis at the aforementioned points to obtain the renormalization coefficients. Our reliance on a free theory motivated approach is justified by the fact that, at high temperatures, QCD interactions become less dominant. Furthermore, this behavior is well-established in the context of the Roberge-Weiss phase transition; since we are only modifying the phase selection criteria, our approach remains theoretically sound.

\section{Result and Analysis}
In this section, we present the results of our methodology and discuss the corresponding findings. As previously outlined, our approach utilizes the measurement of enthalpy density to determine these renormalization coefficients. We extend the findings originally presented in \cite{PN} to additional $\beta$ values at approximately the same temperature; this extension is performed to facilitate a reliable extrapolation to the continuum limit.
We perform our measurements on (1+1+1)-flavor HISQ gauge field configurations with physical strange quark masses and $m_u = m_d = m_s/20$ at $\beta = 7.373$, and $\beta = 7.596$ with volumes $40^3 \times 8 \, (T = 409.7$ MeV) and $48^3 \times 10\, (T = 400.0$ MeV)  respectively. The simulations are carried out for a flavor-dependent imaginary chemical potential configuration defined by $\mu_u / T =  -\mu_d /T = i \theta $ and $\mu_s = 0$. We have performed simulations at seven equally spaced values of $\theta$ within the interval $[0, 2\pi /3]$. \\

\begin{figure}[h]
\centering
\includegraphics[width=0.492\textwidth]{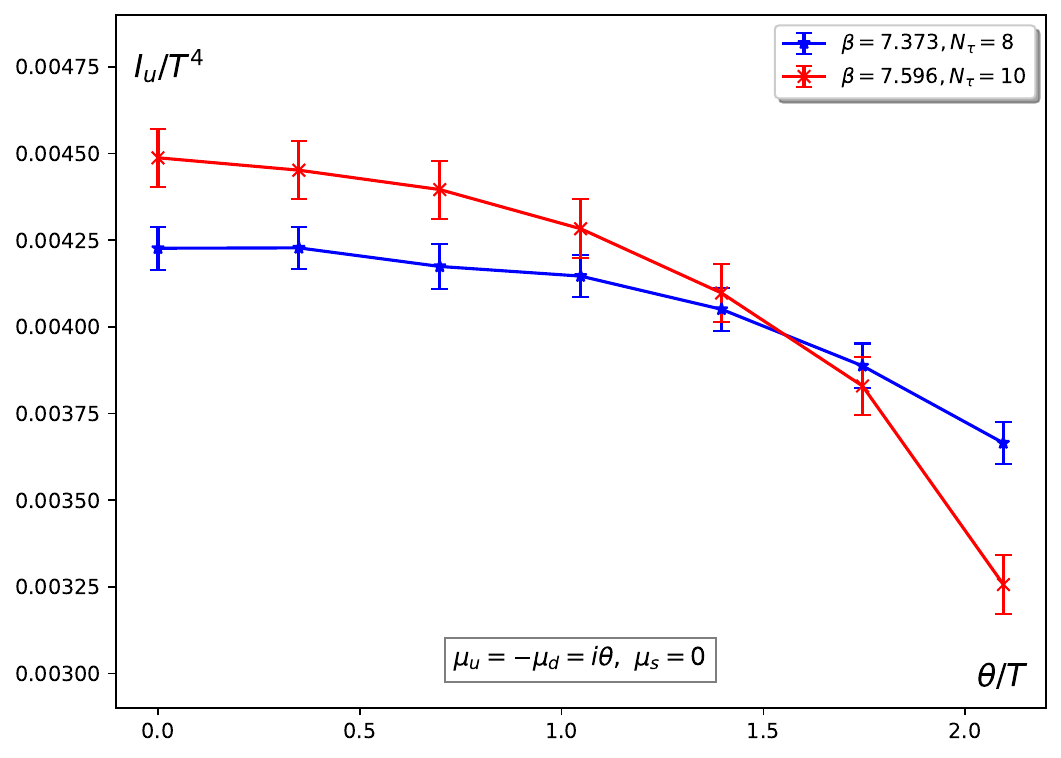}
\includegraphics[width=0.48\textwidth]{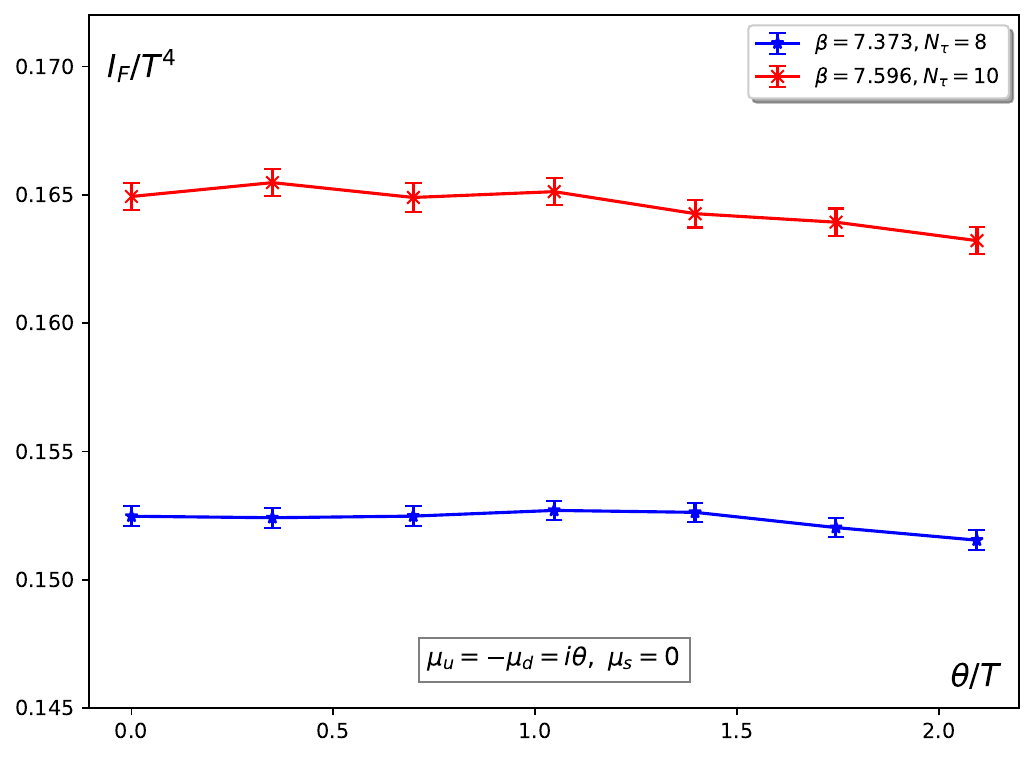}
\caption{The up quark contribution (left) and fermionic contribution (right) to the interaction measure.
\label{fig:3}}
\end{figure}

\begin{figure}[h]
\centering
\includegraphics[width=0.48\textwidth]{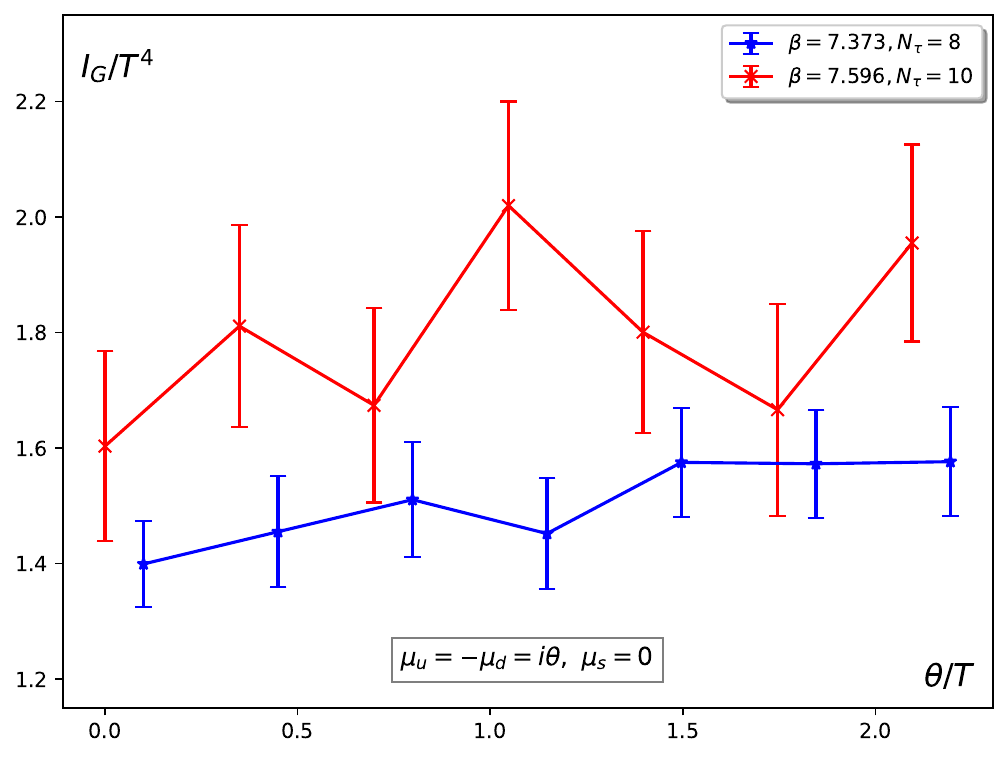}
\includegraphics[width=0.48\textwidth]{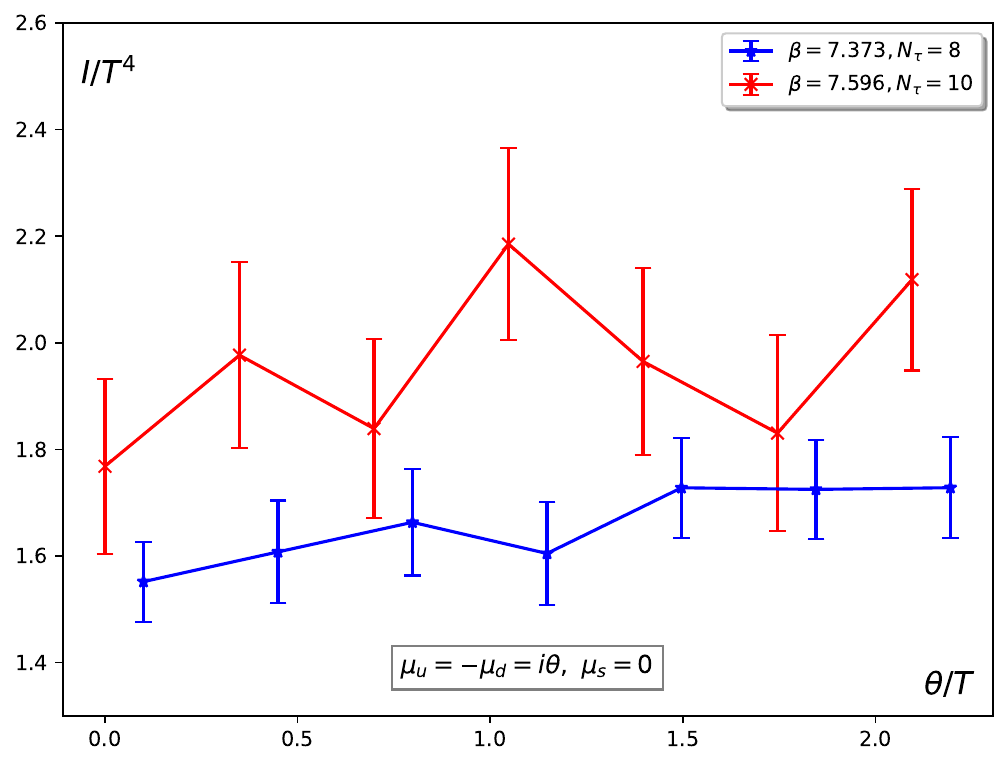}
\caption{The gluonic contribution (left) and overall contribution (right) to interaction measure.
\label{fig:4}}
\end{figure}

To determine the enthalpy density, we need the interaction measure and the pressure.
The interaction measure is calculated using the following expression:
\begin{equation}
\frac{I}{T^4}  
= R_\beta \left [
\left ( \langle S_G \rangle_0 - \langle S_G \rangle_\tau \right ) - R_{m} \sum_{q = u, d, s} 
 m_{q}\left( \langle\bar{\psi}\psi \rangle_{q,0}
- \langle\bar{\psi}\psi \rangle_{q,\tau} \right)
 \right ] N_\tau^4.
 \end{equation}
 The non-perturbative beta function and mass renormalization function are defined in \cite{AB}.
 The zero temperature values for each term are taken from \cite{AB}.
Our results are shown in Fig.~\ref{fig:3} and \ref{fig:4}.

Since pressure is not directly accessible on the lattice, we calculate it by integrating the number density. This necessitates performing measurements at intermediate points, even though our primary analysis focuses on the ensembles at $\theta = 0$ and $\theta = 2 \pi /3$.
The number density and pressure are determined using
\begin{align}
p (\mu_f) & = p(\mu_I = 0) + \int^{\mu_f}_{0} n(\mu) \, d\mu_I
&
n(\mu) & = \frac{1}{4} \frac{T}{V} \, \left\langle \text{Tr} \left[ M(\mu)^{-1} \frac{\partial M(\mu) }{\partial \mu}  \right] \right\rangle
\end{align}
Here $p(\mu_I=0)$ the zero-$\mu$ pressure is taken from data published in \cite{AB} and is $p_0/T^4 = 3.831 \pm 0.134$ ($T = 409.7$ MeV) and $p_0/T^4 = 3.782 \pm 0.132$ ($T = 400.0$ MeV).  

\begin{figure}[H]
\centering
\includegraphics[width=0.489\textwidth]{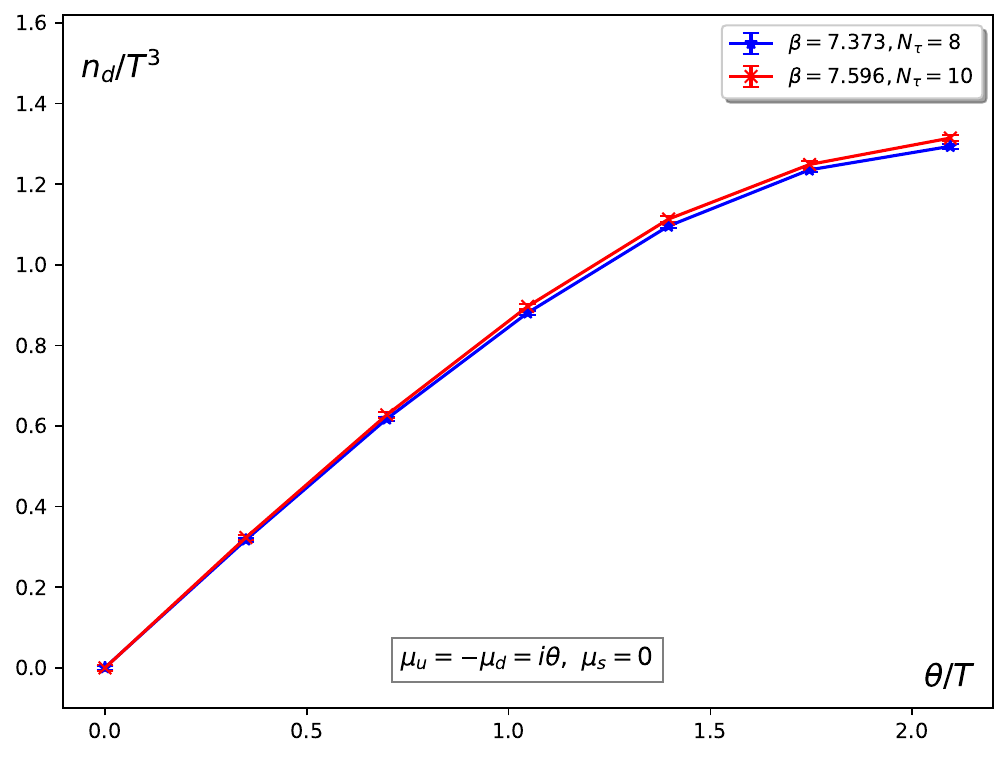}
\includegraphics[width=0.48\textwidth]{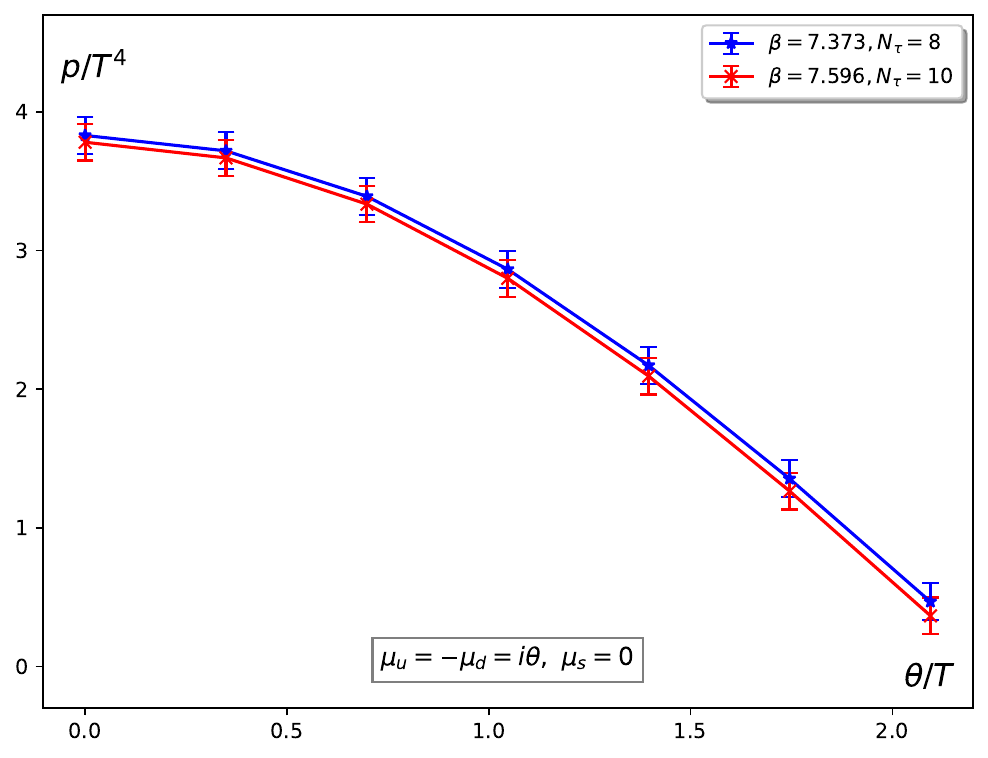}
\caption{The down-quark number density (left) and pressure (right) for $\beta = 7.373$ and  $\beta = 7.596$.
\label{fig:5}}
\end{figure}

\begin{figure}[h]
\centering
\includegraphics[width=0.48\textwidth]{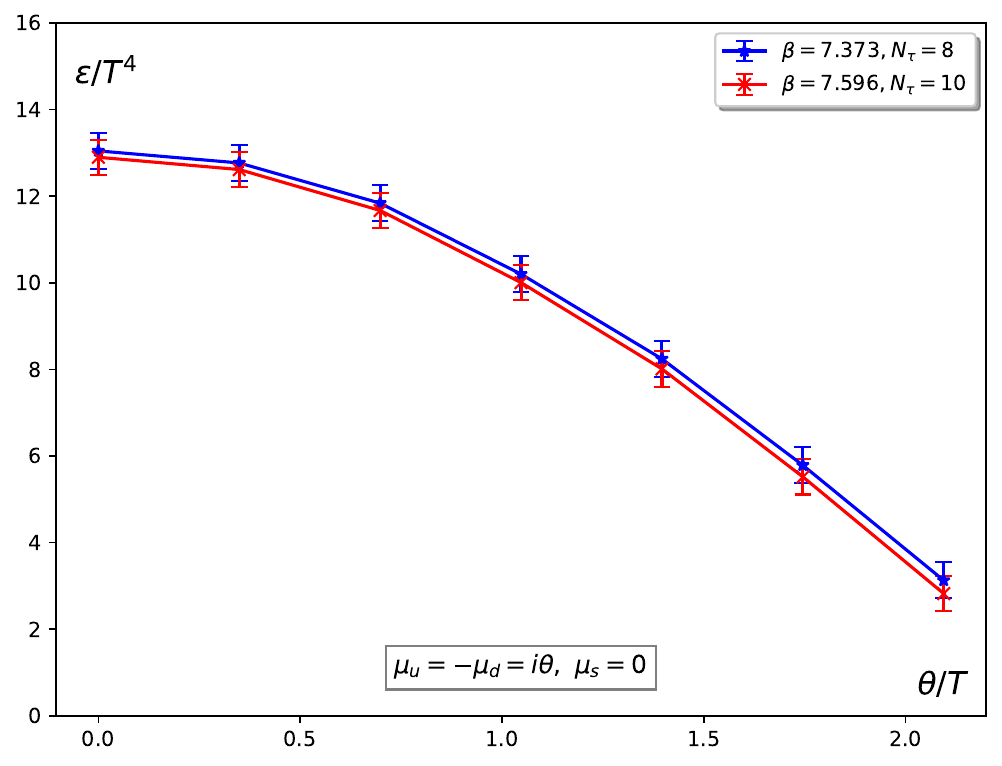}
\includegraphics[width=0.49\textwidth]{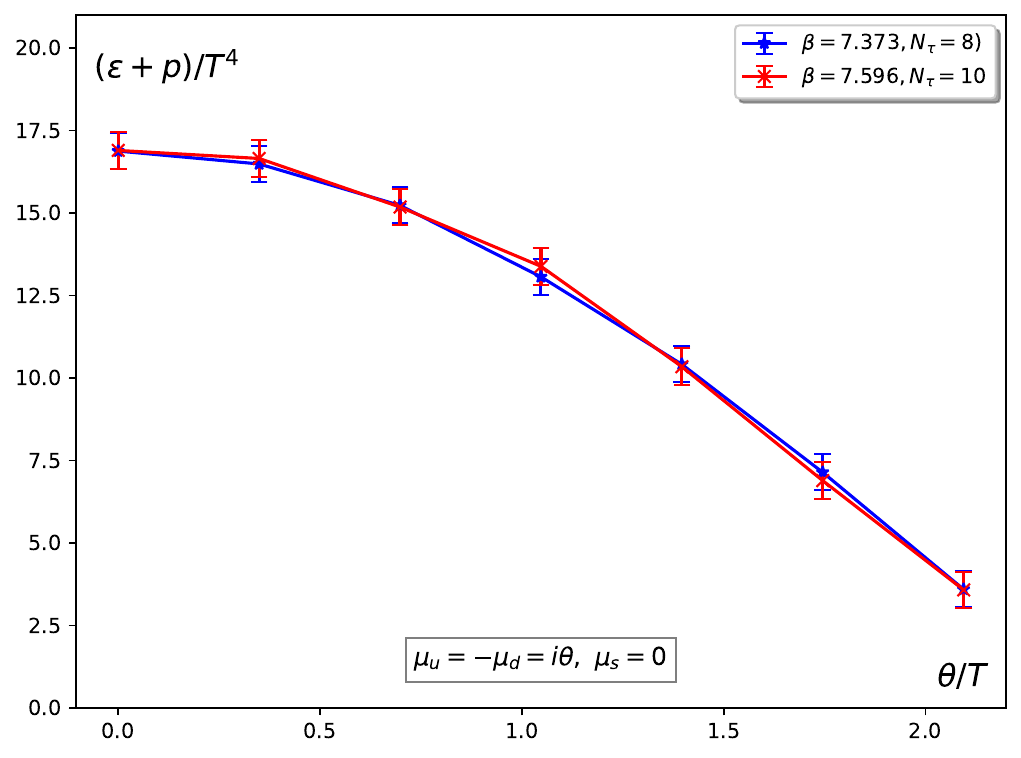}
\caption{The energy density (left) enthalpy density (right) for $\beta = 7.373$ and  $\beta = 7.596$.
\label{fig:6}}
\end{figure}

The number density is measured with high precision; consequently, the uncertainty in the pressure is dominated by zero-$\mu$ pressure. The combinations of $\beta$ and $N_T$ are chosen to ensure that both $\beta = 7.373$ and $\beta = 7.596$ values have identical temperatures, $T = 409.7$ MeV and $T = 400.0$ MeV respectively. As a result, the values for number density, pressure, energy density, and enthalpy density remain highly consistent between the two setups. The smoothness of the observables is also due to the fact that these measurements were performed around $T = 400$ MeV, which is significantly higher than the pseudo-critical temperature ($T_{pc} \sim 155$ MeV).

Fig.~\ref{fig:5} shows as expected that the pressure falls significantly as $\theta \to 2\pi T/3$. But the observed suppression is even stronger than expected: a factor of 10 rather than of 3. While the pressure contribution obtained by integrating the number density is very close to the free-theory value, the gluonic pressure is apparently far below its free value even at $T = 400$ MeV. Thus, the gap between the free and interacting pressure known in the literature
($P_{\text{free}}=5.2T^4$ while $P(T=400\,\text{MeV})=3.78 T^4$ from \cite{AB}) appears to arise from a suppression of gluonic contribution to the pressure and not from the quark contribution.

The enthalpy density in Fig.~\ref{fig:6} is determined without gradient flow, while the EMT operators are measured at finite flow time. This is due to fact that enthalpy density has a plateau region where it is independent of flow-time \cite{MA}.

\section{Conclusion and Outlook}
We have presented an update on the renormalization of the EMT in (2+1)-flavor QCD. Our results, obtained across multiple $\beta$ values at comparable temperatures, demonstrate strong consistency for the number density, pressure, energy density and enthalpy density. To facilitate a reliable continuum extrapolation, we are currently incorporating an additional lattice spacing, $\beta = 7.825$ with a lattice volume of $64^3 \times 12$.
Furthermore, we have shown that applying imaginary isospin chemical potential substantially lowers the fermionic pressure, which validates the robustness of our approach for EMT renormalization.
Our findings also indicate that the gluonic pressure contribution is significantly smaller than the free-theory value at $T = 400$ MeV, whereas the fermionic contribution is nearing to the Stefan-Boltzmann limit.
While our measurements of the number density are highly precise, the primary source of uncertainty remains the zero-$\mu$ pressure contribution.
We are currently refining our analysis to mitigate this error and finalize the renormalization coefficients.
With the EMT operator measurements for the enthalpy density nearing completion, these coefficients will facilitate the first-ever determination of shear viscosity in (2+1)-flavor QCD.

\acknowledgments
We thank Dibyendu Bala, Michele Pepe, and Peter Petreczky for their insightful comments and productive discussions. This work is supported by the Deutsche Forschungsgemeinschaft (DFG, German Research Foundation) via the CRC-TR 211 "Strong-interaction matter under extreme conditions" (Project No. 315477589) and PUNCH4NFDI (Grant No. 460248186). Numerical simulations were conducted on the LUMI-G supercomputer, the GPU cluster at Bielefeld University 
and Noctua 2 at the NHR Center PC2 under the project name hpc-prf-chiral.
We extend our thanks to the Bielefeld HPC.nrw team for their technical assistance. Furthermore, we acknowledge the EuroHPC Joint Undertaking for the allocation of computing time on the LUMI-G system, hosted by CSC (Finland) and the LUMI consortium, via the EuroHPC Extreme Scale Access program.


\begin{thebibliography}{99}
\bibitem{GM}
Guy D. Moore, \emph{Shear viscosity in QCD and why it's hard to calculate}, \href{https://arxiv.org/abs/2010.15704}{\tt{2010.15704}}.

\bibitem{JG}
J. Ghiglieri, G. D. Moore, D. Teaney, \emph{QCD shear viscosity at (almost) NLO}, \href{https://doi.org/10.1007/JHEP03(2018)179}{JHEP \textbf{03} (2018) 179}, [\href{https://arxiv.org/abs/1802.09535}{\tt{1802.09535}}].

\bibitem{LA}
L. Altenkort \textit{et al.}, \emph{Viscosity of pure-glue QCD from the lattice}, \href{https://doi.org/10.1103/PhysRevD.108.014503}{Phys. Rev. D 108 (2023) 014503}, [\href{https://arxiv.org/abs/2211.08230}{\tt{2211.08230}}].

\bibitem{HM}
Harvey B. Meyer, \emph{Calculation of the shear viscosity in SU(3) gluodynamics}, \href{https://doi.org/10.1103/PhysRevD.76.101701}{Phys. Rev. D 76 (2007) 101701(R)}, [\href{https://arxiv.org/abs/0704.1801}{\tt{0704.1801}}].

\bibitem{FW}
F. Karsch and H. W. Wyld, \emph{Thermal Green’s functions and transport coefficients on the lattice}, \href{https://doi.org/10.1103/PhysRevD.35.2518}{Phys. Rev. D 35 (1987) 2518}.

\bibitem{SC}
S. Caracciolo, G. Curci, P. Menotti, A. Pelissetto, \emph{The energy-momentum tensor for lattice gauge theories}, \href{https://doi.org/10.1016/0003-4916(90)90203-Z}{Annals of Physics Volume 197, Issue 1, January 1990, Pages 119-153}.

\bibitem{HS}
Hiroshi Suzuki, \emph{Energy–momentum tensor from the Yang–Mills gradient flow}, \href{https://doi.org/10.1093/ptep/ptt059}{Prog. Theor. Exp. Phys. 2013, 083B03} [\href{https://arxiv.org/abs/1304.0533}{\tt{1304.0533}}].

\bibitem{GP}
L. Giusti and M. Pepe, \emph{Energy-momentum tensor on the lattice: Nonperturbative renormalization in Yang-Mills theory}, \href{https://doi.org/10.1103/PhysRevD.91.114504}{Phys. Rev. D 91 (2015) 114504}, [\href{https://arxiv.org/abs/1503.07042}{\tt{1503.07042}}].

\bibitem{GP2}
M.~Dalla Brida, L.~Giusti and M.~Pepe,
\emph{Non-perturbative definition of the QCD energy-momentum tensor on the lattice,}
\href{https://doi:10.1007/JHEP04(2020)043}{JHEP \textbf{04} (2020), 043}
[\href{https://arxiv.org/abs/2002.06897}{\tt{2002.06897}}].

\bibitem{GP3}
M.~Bresciani, M.~Dalla Brida, L.~Giusti and M.~Pepe,
\emph{Thermal QCD for non-perturbative renormalization of composite operators,}
\href{https://doi:10.22323/1.430.0364}{PoS \textbf{LATTICE2022} (2023), 364}
[\href{https://arxiv.org/abs/2211.13641}{\tt{2211.13641}}].

\bibitem{GP4}
M.~Bresciani, M.~Dalla Brida, L.~Giusti and M.~Pepe,
\emph{Progresses on high-temperature QCD: Equation of State and energy-momentum tensor,}
\href{https://doi:10.22323/1.453.0192}{PoS \textbf{LATTICE2023} (2024), 192}
[\href{https://arxiv.org/abs/2312.11009}{\tt{2312.11009}}].

\bibitem{ML}
Martin Lüscher, \emph{Properties and uses of the Wilson flow in lattice QCD}, \href{https://link.springer.com/article/10.1007/JHEP08(2010)071}{JHEP \textbf{08} (2010), 071} [\href{https://arxiv.org/abs/1006.4518}{\tt{1006.4518}}].

\bibitem{HH}
H. Makino and H. Suzuki, \emph{Lattice energy–momentum tensor from the Yang–Mills gradient flow—inclusion of fermion
fields}, \href{https://doi.org/10.1093/ptep/ptu070}{PTEP (2014) 063B02}, [\href{https://arxiv.org/abs/1403.4772}{\tt{1403.4772}}], [Erratum: \href{https://doi.org/10.1093/ptep/ptv095}{PTEP (2015) 079202}].

\bibitem{RW}
A.  Roberge  and  N.  Weiss, \emph{Gauge  Theories  With  Imaginary  Chemical  Potential  and  the Phases of QCD}, \href{https://doi.org/10.1016/0550-3213(86)90582-1}{Nucl. Phys. B \textbf{275} (1986) 734}.

\bibitem{PN}
Pavan, O. Kaczmarek, G. D. Moore, C. Schmidt,
\emph{Shear viscosity from quenched to full lattice QCD,}
\href{https://doi.org/10.22323/1.466.0199}{PoS \textbf{LATTICE2024} (2025), 199}
[\href{https://https://arxiv.org/abs/2503.11395}{\tt{2503.11395}}].

\bibitem{AB}
A. Bazavov \textit{et al.}, \emph{Equation of state in (2+1)-flavor QCD}, \href{https://doi.org/10.1103/PhysRevD.90.094503}{Phys. Rev. D 90 (2014) 094503}, [\href{https://arxiv.org/abs/1407.6387}{\tt{1407.6387}}].

\bibitem{MA}
M. Asakawa \textit{et al.}, \emph{Thermodynamics of $SU(3)$
gauge theory from gradient flow on the lattice}, \href{https://doi.org/10.1103/PhysRevD.90.011501}{Phys. Rev. D 90 011501(R)}, [\href{https://https://arxiv.org/abs/1312.7492}{\tt{1312.7492}}].


\end{thebibliography}
\end{document}